\documentclass[aps,pra,twocolumn,amsmath,nofootinbib]{revtex4-1}
\usepackage{epsfig}
\usepackage[colorlinks,linkcolor=blue,anchorcolor=blue,
            citecolor=blue,urlcolor=blue,
            breaklinks=true]{hyperref}
\usepackage{graphics}
\usepackage{color}
\usepackage{graphicx}
\usepackage{dcolumn}
\usepackage{bm}
\usepackage{cases}
\usepackage{amsfonts}
\usepackage{appendix}
\usepackage{ulem}

\begin{document}
\author{Xin-Sheng Zhang$^{1}$}
\author{Wen-Jie Xu$^{1}$}
\author{Hang-Yu Gu$^{1}$}
\author{Chuan-Xun Du$^{1,3}$}
\email{duchuanxun@lyu.edu.cn}
\author{Yong-Long Wang$^{1,2,3}$}
\email{wangyonglong@lyu.edu.cn}

\address{$^{1}$ School of Physics and Electronic Engineering, Linyi University, Linyi, 276000, China}
\address{$^{2}$ Department of Physics, Nanjing University, Nanjing 210093, China}
\address{$^{3}$ National Laboratory of Solid State Microstructures, Department of Materials Science and Engineering, Nanjing University, Nanjing, 210093, China}

\title{Realizing Next-Nearest-Neighbor Coupling and Peierls Phase in Circuits}
\begin{abstract}
We really design the trimerized circuits for the non-Hermitian one-dimensional Su-Schrieffer-Heeger models. There are three models, the initial one just considers the nearest neighbor coupling, the enhanced one is extended to contain the next-nearest-neighbor coupling, and the final one is reenhanced by introducing the Peierls phase. We investigate the dynamics of the circuit Laplacians with respect to the models, find that the topological states appear in the initial model and the response intervals are substantially affected by the next-nearest-neighbor coupling channels and the Peierls phase. These results are practically demonstrated by numerical simulations and experimental measurements. As a conclusion, the trimerized circuits can provide an adjustable and simple platform to investigate new topological physical states.
\bigskip

\noindent PACS Numbers: 84.30.-r, 03.67.Ac, 03.65.-w, 03.65.Vf
\end{abstract}
\maketitle

\section{Introduction}\label{1}
In discrete lattice systems, next-nearest-neighbor (NNN) coupling and the Peierls phase constitute two key ingredients for extending coupling  access and introducing gauge-phase modulation. The  introduced NNN coupling term  goes beyond a simple correction to the nearest-neighbor (NN) coupling approximation, and it provides a more complete description of band structure and topological-phase formation, as demonstrated in two-dimensional Dirac lattices\cite{Beugeling2012PRB}, Su-Schrieffer-Heeger (SSH) lattice\cite{Geng2022NatCommun}, generalized SSH models\cite{Li2014PRB}, long-range SSH models\cite{Hsu2020PRB}, and parity-time (PT)-symmetric trimerized lattices\cite{Du2021OptExpress}. Moreover, in the Haldane model\cite{Maiti2019PRB, Guo2025CPB}, Heisenberg model\cite{Yao2022PRB} and Kane-Mele lattices\cite{Morad2022PRB}, the NNN  coupling plays a structural role in gap opening, symmetry breaking, and the emergence of new phases\cite{Zvyagin2001EPJB} and nontrivial boundary states. In other words, the NNN interaction can provide a new access to enhance the controllability of the previously introduced systems.

Moreover, the Peierls phase can provide the other access to reconstruct and manipulate new topological states\cite{Gotz2024PRB, Zhang2026PLA} in the considered systems. Specifically, in the crossover fields of the quantum information technology and the material science\cite{Luo2023PRB, Niveth2025JPCM}, the Peierls phase was employed to produce higher-order topological phases and nontrivial insulator. Physically, the Peierls phase plays the essential role of emergent gauge field\cite{Gorg2019NatPhys, Ohler2022NJP}. In the framework, the physically relevant quantity is not only an isolated phase factor on a single bond, but also the gauge-invariant phase accumulated around a closed loop. When the NNN channels participate in the formation of closed coupling paths, such loop-phase accumulation modifies the interference among different propagation paths and thereby provides a natural route to reconfigurable spectral and topological control\cite{Islam2022PRB}. To further investigate the effects of the NNN coupling and the Peierls phase, the photonic systems\cite{Ozawa2019RMP}, waveguide arrays and atomic lattices\cite{Drost2017NatPhys} are experimentally employed to be platforms. In those experimental systems, it is the restriction of applications that the parameters of the NNN coupling and the Peierls phase are in general difficult to change. An alternative platform with more adjustable parameters needs to be investigated the actions of the NNN coupling and the Peierls phase.

A good alternative candidate, the circuit consisting of resistor, inductor and capacitor (RLC) was recently employed to investigate the topological states\cite{Lee2018CommunPhys, Imhof2018NatPhys, Zhang2019PRB, Yan2024PhysRep, Sahin2025APL}. Initially, the behavior of an RLC circuit is governed by its circuit Laplacian, which is analogous to the Hamiltonian describing the energy band of a physical quantum system\cite{Lee2018CommunPhys}. Strikingly, the abundant elements and flexible connections enable the RLC circuits to provide more convenient accesses to construct and manipulate the topological states\cite{Yan2024PhysRep, Sahin2025APL}, such as coupling strength, coupling phase, gain and loss, synthetic dimensions, artificial gauge field and so on\cite{Kollar2024NatPhys}.  Potentially, the topological circuits are also compatible with integrated-circuit technology, which is advantageous for manufacture, miniaturization, and prospective device applications\cite{Sahin2025APL, Ji2025PRB, Zhang2026PRB}. In other words, the further investigations on realizing the NNN coupling and Peierls phase in circuits are of scientific and applicable values.

In the present paper, we will try to experimentally design a topological circuit, in which the NNN coupling and the Peierls phase are introduced by connections to modify the topological states. And the paper is organized as follows. In Sec.\ref{2}, according to the known circuit for the non-Hermitian one-dimensional SSH lattice, we first design the non-Hermitian one-dimensional trimer circuit with the NNN coupling and the Peierls phase, and deduce the effective Hamiltonian with respect to the designed circuits. In Sec.\ref{3}, we compare the initial, enhanced by NNN coupling, and reenhanced by Peierls phase models to show how the added NNN coupling channel regulates the topological states and how the Peierls phase redistributes the stability response intervals and localization strengths of the upper and lower edge states. In Sec.\ref{4}, numerical simulations and experimental measurements are performed to verify the predicted phase-dependent shifts of the edge-state response intervals and the modulation of boundary-voltage localization. In Sec.\ref{5}, some conclusions and discussions are given.

\section{Realisations in Circuits for Su-Schrieffer-Heeger Models}\label{2}
\begin{figure}[htbp]
	\includegraphics[width=\linewidth]{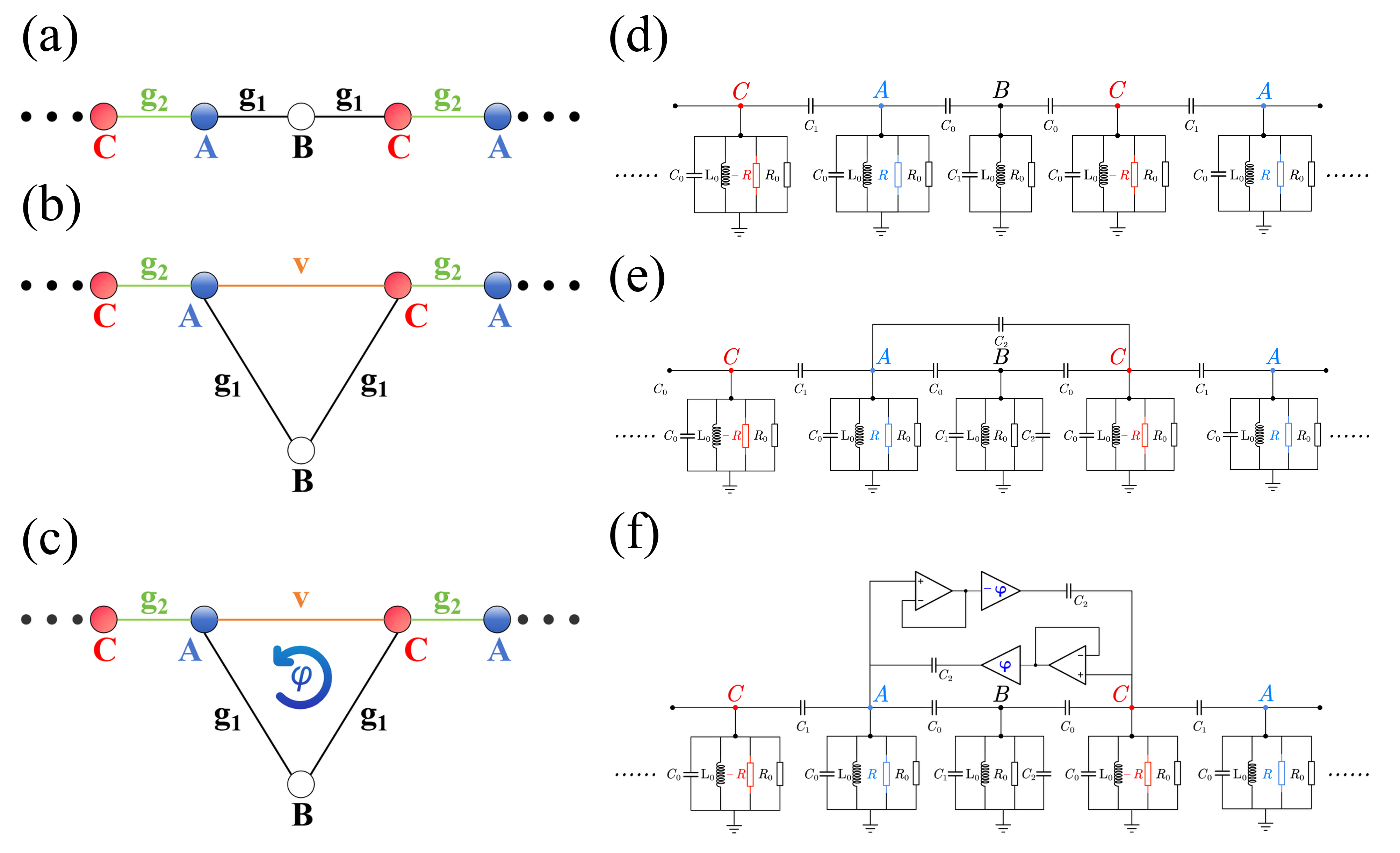}
	\caption{Theoretical models and circuit implementations of the non-Hermitian one-dimensional trimerized circuits. (a) Initially trimerized SSH model with NN couplings $g_1$ and $g_2$. (b) Enhanced trimerized SSH model with an additional NNN coupling $\nu$. (c) Reenhanced trimerized SSH model with an NNN coupling $\nu$ and a Peierls phase $\varphi$. (d)--(f) Circuit implementations corresponding to panels (a)--(c), respectively.}\label{Fig1}
\end{figure}
 In electrical engineering, the non-Hermitian one-dimensional trimerized SSH model just with the NN coupling can be first realized by a topological circuit, which consists of the networks of resistors, inductors and capacitors, as shown in Figs.\ref{Fig1}(a) and (d). In the initial model, the lattices are periodically constructed by trimerized unit cells. Each unit cell contains three sublattice sites, they are in turn a loss node labeled as $A$, a neutral one as $B$ and a gain site as $C$.  The lattice node $B$ is grounding through an RLC circuit consisting of a resistor $R_0$, an inductor $L_0$ and a capacitor $C_1$. In contrast with the node $B$, the capacitor $C_1$ is replaced by $C_0$, and the grounding RLC circuit is additionally introduced a positive resistor $R$ for the node $A$, a negative resistor $-R$ for the node $C$. The nodes $B$ are connected to $A$ and $C$ with a same capacitor $C_0$ that plays the role of coupling labeled as $g_1$. Differently, the nodes $A$ and $C$ are connected by another capacitor $C_1$, the corresponding coupling is labeled as $g_2$.  The difference between the couplings $g_1$ and $g_2$ mainly determines the topological response in the considered model, and the loss of node $A$ and the gain of node $C$ together give rise to non-Hermitian.

 For the initial SSH circuit sketched in Fig.\ref{Fig1} (a),  the node equations  can be  described by the circuit-Laplacian in the following equation\cite{Lee2018CommunPhys}
\begin{equation}\label{Laplacian}
	\begin{aligned}
		I_a^0=\sum_b J_{ab}^0(\omega)V_b^0,
	\end{aligned}
\end{equation}
where $I_a^0$ is the external current injected into the node $a$, $V_b^0$ is the voltage at node $b$, and $J_{ab}^0(\omega)$ is the circuit Laplacian, $\omega$ is the frequency with $e^{i\omega t}$, and $a, b=A, B, C$ denote an arbitrary node in the circuit Fig.\ref{Fig1}(a).  Generally, the Laplacian can be expressed as
$J^0_{ab}(\omega)=i\omega C_{ab}(\omega)+\sigma_{ab}+1/[i\omega L_{ab}(\omega)]$,
 wherein $C_{ab}$ denotes the capacitance between  the nodes $a$ and $b$, $\sigma_{ab}$  stands for the conductance between $a$ and $b$  and $L_{ab}$ describes the inductance between $a$ and $b$.  For the initial SSH circuit Fig.\ref{Fig1} (d), the circuit Laplacian  is specifically given by
\begin{equation}\label{Laplacian0}
	\begin{aligned}
		J^0(\omega,k) &= i\omega \left(2C_0 + C_1 - \frac{1}{\omega^2 L_0} \right)\mathbb{E} \\
		&\quad + i\omega
		\begin{pmatrix}
			\frac{\sigma+\sigma_0}{i\omega} & -C_0 & -C_1 e^{-ik} \\
			-C_0 & \frac{\sigma_0}{i\omega} & -C_0 \\
			-C_1 e^{ik} & -C_0 & \frac{-\sigma+\sigma_0}{i\omega}
		\end{pmatrix},
	\end{aligned}
\end{equation}
where $\mathbb{E}$ is the identity matrix, and $k$ is the Bloch wave number introduced  by the periodic boundary condition $V_{n\pm1}(t)=V_n(t)e^{\pm ik}$, the terms in the parenthesis before $\mathbb{E}$ are contributed by two capacitors $C_0$, one capacitor $C_1$ and one inductor $L_0$ connecting the nodes, in the diagonal elements the conductances $\sigma$, $-\sigma$ and $\sigma_0$ are provided by the resistors $R$, $-R$ and $R_0$, respectively, and the nondiagonal elements are given by the couplings between the nodes $A$, $B$ and $C$. Obviously, the non-Hermiticity of the circuit Laplacian $J^0$ is determined by the positive conductance $\sigma$ and the negative one $-\sigma$ without $\sigma_0$, and the Laplacian $J^0$ is a function of $\omega$.  Nevertheless, the Laplacian $J^0$ can be employed to analyze the spectrum and topological invariants\cite{Liu2020PRAppl}.  As the frequency takes $\omega_0=\frac{1}{\sqrt{(2C_0+C_1)L_0}}$, the inductive and capacitive diagonal contributions cancel each other. In other words, the terms before $\mathbb{E}$ in Eq.\eqref{Laplacian0} are vanished.  And then normalizing by $C_0$, the dimensionless effective circuit Laplacian $J^0_n$ can be simplified in the following form
\begin{equation}\label{NormLaplacian0}
	J_{n}^0(\omega_0,k)=
	\begin{pmatrix}
		-i\gamma & -g_1 & -g_2e^{-ik}\\
		-g_1 & 0 & -g_1\\
		-g_2e^{ik} & -g_1 & i\gamma
	\end{pmatrix},
\end{equation}
where $g_1=1$ describes the intracell coupling strength, $g_2=C_1/C_0$ does the intercell one, and $\gamma=\sigma\sqrt{L_0/C_0}\sqrt{2+g_2}$ stands for the non-Hermitian gain-loss strength.

According to the node equations Eq.\eqref{Laplacian}, from the circuit Laplacian $J^0_{ab}(\omega)$ we can obtain the effective Hamiltonian as
(see Appendix~\ref{A} and Eqs.\eqref{SecondKEq}-\eqref{EffHam}),
\begin{equation}
	H=i
	\begin{pmatrix}
		C^{-1}\sigma & C^{-1}L^{-1} \\
		-\mathbb{E} & 0
	\end{pmatrix}.
\end{equation}
In the calculation process, the circuit Laplacian $J_{ab}^0=i\omega C_{ab}+\sigma_{ab}+1/[i\omega L_{ab}]$ and $\partial/\partial t\to i\omega$ are considered. In other words, the effective Hamiltonian $H$ is from the circuit Laplacian $J^0_{ab}$ by multiplying a constant and equivalently replacing. Therefore the topological structures implied in the effective Hamiltonian agree well with those in the circuit Laplacian\cite{Hofmann2019PRL}. Specifically, the eigenvalues of the effective Hamiltonian, namely the circuit eigenfrequencies, equivalently play the role of the eigenvalue points with respect to the edge states in Laplacian spectrum\cite{Lee2018CommunPhys,Hofmann2019PRL,Imhof2018NatPhys}. This equivalence allows the same circuit design to be discussed either as a response network or as an effective non-Hermitian lattice Hamiltonian.

Based on the initial trimerized SSH circuit, we can enhance it by introducing an NNN coupling  by connecting the loss node $A$ and the gain node $C$ with an additional capacitor $C_2$, as shown in Figs.\ref{Fig1}(b) and (e). In the  enhanced model, the additional coupling channel is denoted by $\nu$,  which links the $A$ and $C$ sublattices within the same trimerized unit cell. In the  enhanced circuit, the NNN coupling is realized by the coupling capacitor $C_2$,  that is the coupling strength $\nu=C_2/C_0$ normalized by $C_0$. The capacitor $C_2$ therefore provides an additional hopping  between the $A$ and $C$ nodes, extending the  initial NN chain to an NNN-coupled trimerized chain and forming the closed coupling channel  with phase accumulation. By introducing the NNN coupling, there appears an additional connection between the loss node $A$ and the gain node $C$, and there is a parameter $-\nu$ added to the elements of $J^0_{n,13}$ and $J^0_{n,31}$. And the effective enhanced and normalized circuit Laplacian $J^e_n$ can be simplified as
\begin{equation}\label{LaplacianNNN}
	J^e_n(\omega_0,k)=
	\begin{pmatrix}
		-i\gamma & -g_1 & -g_2 e^{-ik}-\nu \\
		-g_1 & 0 & -g_1 \\
		-g_2 e^{ik}-\nu & -g_1 & i\gamma
	\end{pmatrix},
\end{equation}
where $\nu$ is  the strength of the NNN coupling channel, which can tune the relative weight of the direct $A$--$C$ path against the NN-mediated path.

Further, the enhanced SSH circuit can be reenhanced again by introducing a Peierls phase through a nonreciprocal phase-shift circuit introduced into the NNN coupling channel\cite{Albert2015PRL}, as shown in Figs.\ref{Fig1}(c) and (f).  The module imposes a controllable phase delay on the coupling  between the nodes $A$ and $C$ within the same unit cell, so that the NNN coupling acquires direction-dependent complex phases while its amplitude remains unchanged\cite{Peierls1933ZPhys,Hofstadter1976PRB}.
The Peierls phase is therefore incorporated directly into the complex coupling term rather than into the NN coupling. With the Peierls phase  and the NNN coupling, the effective reenhanced and normalized circuit Laplacian $J^{re}_n(\omega_0, k)$ can be given by
\begin{equation}\label{LaplacianPP}
	J^{re}_n(\omega_0,k)=
	\begin{pmatrix}
		-i\gamma & -g_1 & -g_2 e^{-ik}-\nu e^{i\varphi} \\
		-g_1 & 0 & -g_1 \\
		-g_2 e^{ik}-\nu e^{-i\varphi} & -g_1 & i\gamma
	\end{pmatrix},
\end{equation}
where $\varphi$ is the Peierls phase introduced into the NNN coupling channel. When $\varphi=0$, Eq.\eqref{LaplacianPP} reduces to the effective enhanced and normalized circuit Laplacian $J^e_n(\omega_0, k)$. As $\nu=0$, it would further reduce to the initial circuit Laplacian $J^0_n(\omega_0, k)$. Thus, the Peierls phase does not change the NN coupling structure and the NNN coupling amplitude.  Remarkably, it introduces a controllable phase degree of freedom into the closed coupling loop,  and provides an additional access to investigate the topological response.

\section{Topological States in the Non-Hermitian SSH Circuit}\label{3}
In this section, we will first numerically investigate the spectrum for the three SSH circuits consisting of $8$ unit cells, such as the initial SSH circuit, the SSH circuit enhanced by the NNN coupling channels and the SSH circuit reenhanced by the Peierls phase again. For simplicity, in these calculations we will take $n=8$, $g_1=1$, $\gamma=0.2$ and $g_2=g_1-\Delta\cos\theta$, where $\Delta=0.5$ and $\theta\in[0, 2\pi]$ in Eqs.\eqref{NormLaplacian0}, \eqref{LaplacianNNN} and \eqref{LaplacianPP}. The circuit Laplacian $J^0_n$ is that $J^{re}_n$ of $\nu=0$ and $\varphi=0$, $J^{e}_n$ is $J^{re}_n$ of $\varphi=0$. The three circuit Laplacians can be taken as the effective Hamiltonians, and can be employed to calculate the effective description spectra in the present paper.

\subsection{Topological State Evolution Induced by NNN Coupling}\label{3S1}
\begin{figure}[htbp]
	\includegraphics[width=\columnwidth]{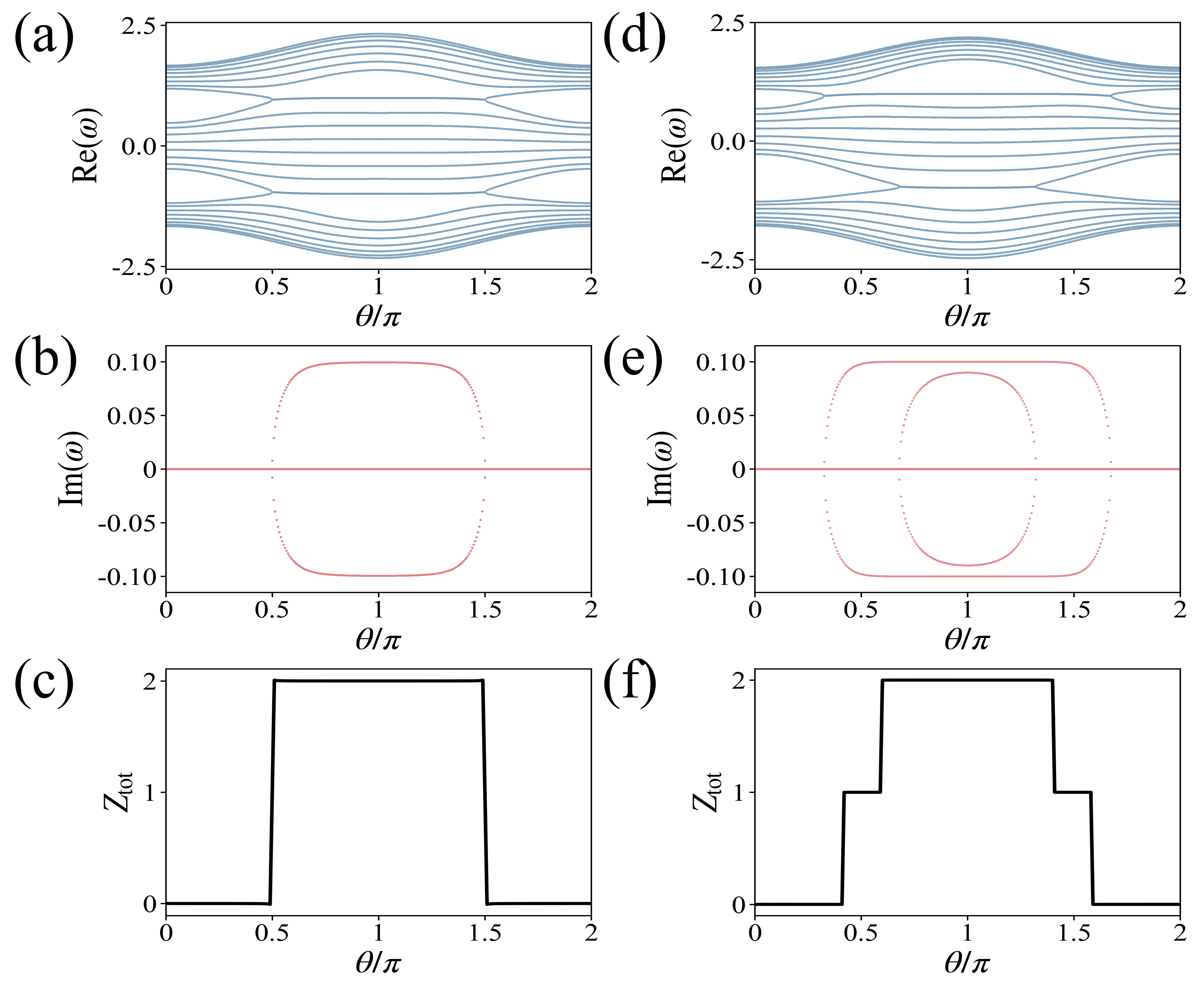}
	\caption{Eigenfrequency spectra and total Zak phase for the  initial and NNN-coupled models showing the transition from a two-sector response to a stepwise topological sequence. (a) The real parts of the eigenfrequency spectra for $\nu=0$ and $\varphi=0$; (b) The imaginary parts for $\nu=0$ and $\varphi=0$; (c) The total Zak phase $z_{tot}$ of $\nu=0$ and $\varphi=0$; (d) The real parts for $\nu=0.2$ and $\varphi=0$; (e) The imaginary parts for $\nu=0.2$ and $\varphi=0$; and (f) The total Zak phase $z_{tot}$ of $\nu=0.2$ and $\varphi=0$.}
	\label{Fig2}
\end{figure}
 The initial non-Hermitian one-dimensional trimer model  with $\nu=0$ and $\varphi=0$ is described by $J^0_n(\omega_0, k)$, which is taken as an effective Hamiltonian that can be calculated by the secular equation. In terms of $J^0_n(\omega_0, k)$, the eigenfrequencies are obtained, and the real and imaginary parts are sketched in Figs.\ref{Fig2}(a) and (c), respectively. In the real-part spectra, the three branches merge within $\theta\in[0.5\pi,1.5\pi]$, and the width of this merging interval is  induced by the nonvanishing of the non-Hermitian parameter $\gamma$, and becomes wider with the increasing of $\gamma$. Corresponding to the merging region of the real part of the eigenfrequencies $\omega$, the imaginary part has three values for each value of $\theta$. It indicates that a PT symmetry-breaking phase transition, the state gain with respect to ${\rm{Im}}(\omega)<0$ and the state loss with respect to ${\rm{Im}}(\omega)>0$. Generally, the bifurcation point of ${\rm{Im}}(\omega)$ is an exceptional point, at which the eigenfrequencies and eigenstates (that is the eigenvectors) are simultaneously degenerate. In other words, in the merging region $\theta\in[0.5\pi, 1.5\pi]$ the voltage with respect to the eigenvalue $E_{8}$ agrees well with that of $E_{9}$, and that of $E_{16}$ is completely equivalent to that of $E_{17}$, where $8$, $9$, $16$ and $17$ denote the energy-like spectra. The interesting results are from the nontrivial topology of the edge states of the initial SSH circuit. Since the  initial model preserves chiral symmetry, its topological origin of these edge states can be characterized by the total Zak phase\cite{Du2021OptExpress,Zak1989PRL,Kunst2018PRL},
\begin{equation}\label{ZakPhase}
z_{tot}=\frac{-i}{2\pi}\sum_{\alpha}\int_{0}^{2\pi}\langle\psi_{m,\alpha}^{\prime}\mid\partial_{k}\psi_{m,\alpha}^{\prime}\rangle dk,	
\end{equation}
where $\alpha \in \{1,2,3\}$ labels the three eigenvalues $E'_{m,\alpha}$ of the effective Laplacian  $J^0_n(\omega_0, k)$,
 which are obtained after diagonalization in the subspace spanned by the sites $B$ and $C$.
The states $\langle \psi_{m,\alpha}^{\prime}|$ and $|\psi_{m,\alpha}^{\prime}\rangle$ are the corresponding left and right eigenstates, respectively.
As shown in Fig.\ref{Fig2}(b), it is strikingly that there appears a topologically nontrivial region with $z_{tot}=2$, for $|\theta-\pi|<0.5\pi$, a trivial region with $z_{tot}=0$ for $|\theta-\pi|>0.5\pi$. However, there are two topologically regions with $z_{tot}=1, 2$ in Fig.\ref{Fig2}(e).

\begin{figure}[htbp]
	\includegraphics[width=\columnwidth]{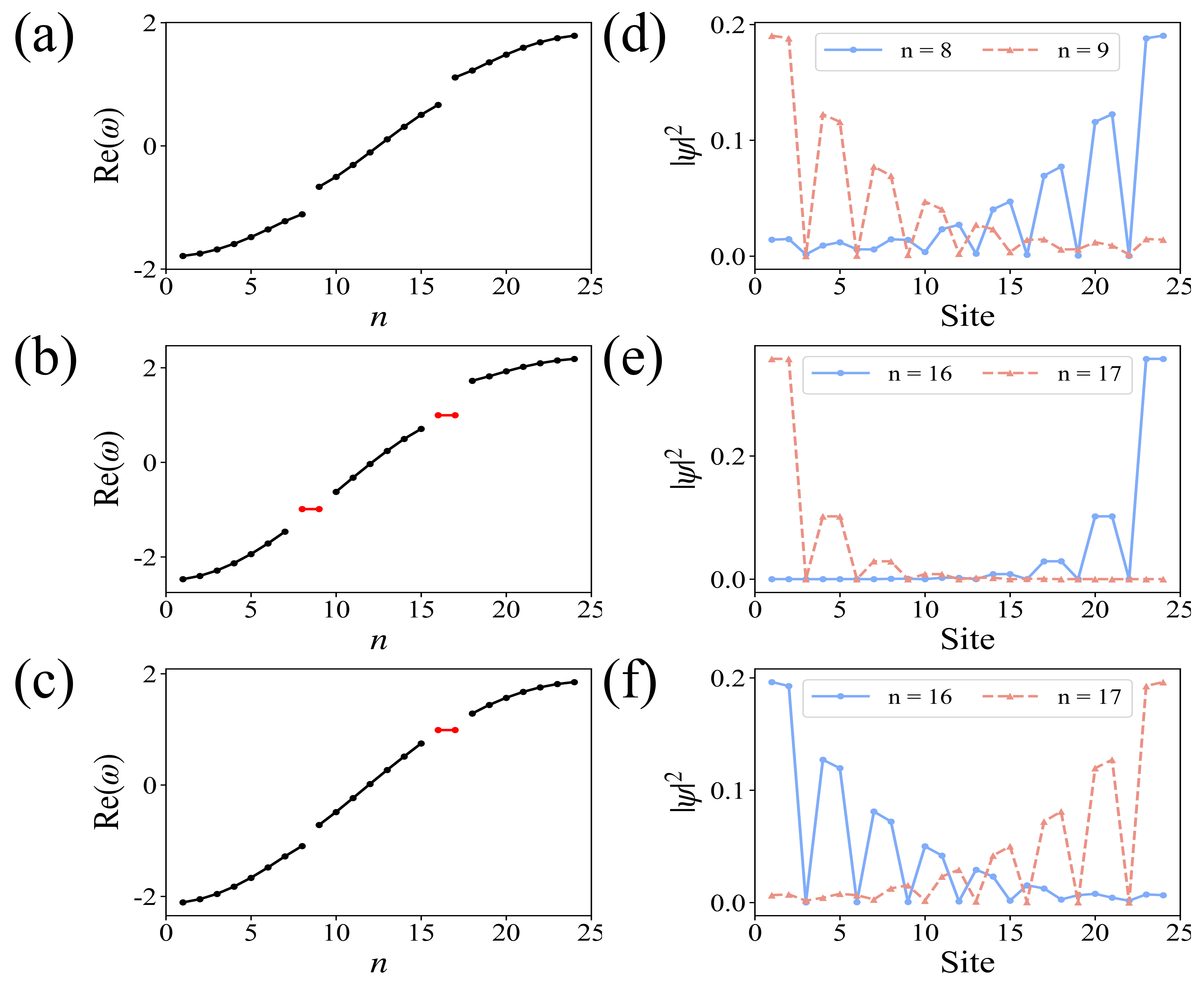}
	\caption{Real-part eigenfrequency spectra and corresponding eigenstate distributions at representative parameters used to identify the boundary-localized sectors. (a)--(c) Real parts of the eigenfrequency spectra for $\theta=0.3\pi$, $\nu=0$; $\theta=\pi$, $\nu=0$; and $\theta=0.5\pi$, $\nu=0.2$, $\varphi=0$, respectively. (d), (e) Spatial distributions of the eigenstates associated with the red isolated eigenvalues in panel (b). (f) Spatial distribution of the eigenstate associated with the red isolated eigenvalue in panel (c). Black dots denote bulk-state branches, and red dots denote isolated edge states separated from the bulk bands.}
	\label{Fig3}
\end{figure}

In order to investigate the specific topological response for the nonvanishing imaginary part of the eigenfrequencies ${\rm{Im}}(\omega)$, the real part of the eigenfrequencies ${\rm{Re}}(\omega)$ changing over the sites $n$ from $1$ to $24$ is described in Fig.\ref{Fig3}. As shown in Fig.\ref{Fig3}(a), there are only bulk states without any edge states for $\theta=0.3\pi$ and $\nu=0$. However, Fig.\ref{Fig3}(b) shows that there are two pairs of edge-states for $\theta=\pi$ and $\nu=0$, the eigenstate of the site of $A_{8}$ is degenerate with that of $B_1$, and the eigenstate of $B_{8}$ is degenerate with that of $C_1$, too. The eigenstates of $E_{8}$ and $E_{9}$ are described in Fig.\ref{Fig3}(d), and those of $E_{16}$ and $E_{17}$ are sketched in Fig.\ref{Fig3}(e). Obviously, in the region of $\theta\in[0.5\pi, 1.5\pi]$ the Zak phase takes value $2$ shown in Fig.\ref{Fig2}(e), and the two pairs of degenerate eigenstates have exactly the same probability distribution as shown in Figs.\ref{Fig3}(d) and (e). The probability distributions mean that the eigenstates in the initial SSH circuit are linearly superposition states, not pure states.

In the enhanced SSH circuit for $\theta=0.5\pi$ and $\nu=0.2$, the states of $E_{8}$ and $E_{16}$ sites in the initial unit cell merge with those of $E_{9}$ and $E_{17}$ sites in the end unit cell with different intervals for $\theta$, respectively. Noticeably, the nonvanishing NNN coupling modifies the Zak-phase taking more values $0\to 1\to 2\to 1\to 0$ in the interval $\theta\in[0,2\pi]$, shown in Fig.\ref{Fig2}(f). Specifically, as $z_{tot}=1$ there is only a pair of degenerate eigenstates of $E_{16}$ and $E_{17}$, when $z_{tot}=2$ there are two pairs of degenerate eigenstates they agree well with the initial SSH circuit. And the corresponding eigenstate distributions are described in Fig.\ref{Fig3}(f).  In other words, the introduced NNN coupling can modify the merging interval of $\theta$ for the different pairs degenerate eigenstates.

\subsection{Topological State Evolution Induced by the Peierls Phase}\label{3S2}
 In the subsection, the enhanced SSH circuit is reenhanced by introducing a Peierls phase in the NNN coupling, that is $\nu\neq 0$ and $\varphi\neq 0$. The presence of phase does not change the amplitude of the NNN coupling $\nu$, but it can effectively modify the momentum-like $\theta$ values, which is specifically described by adding $-\nu e^{i\varphi}$ to $-g_2 e^{-ik}$ to further modify $g_2=g_1-\Delta\cos\theta$. The Peierls phase plays the role of synthetic gauge fields in the reenhanced SSH circuit, which can regulate the topological response in the interval of $\theta$. Numerically, we will investigate the topological nontrivial phase regions $\Delta\theta_{\pm}$ and their localization strengths $\mathrm{IPRs}_{\pm}$ as $\varphi$ functions.

Without loss of generality, we focus on edge solutions localized at the right boundary.  For the simplicity of sake, the loss point can be taken as a vanishing reference. As a consequence, there are two conditions $a_n=0$ and $|c_{n-1}^{\pm}|<|c_n^{\pm}|$. The second condition ensures that the eigenstate decays from the boundary into the bulk. From $a_n=0$, we obtain
\begin{equation}
	\begin{split}
		& 0=g_{1}b_{n}+g_{2}c_{n-1}+\nu e^{i\varphi}c_{n}, \\
		& E_{r}b_{n}=-g_{1}c_{n}, \\
		& (E_{r}-i\gamma)c_{n}=-g_{1}b_{n}.
	\end{split}%
\end{equation}%
These relations determine the eigenvalues of the upper and lower edge states and the amplitude ratio between neighboring unit cells,
\begin{equation}\label{Edge}
	\begin{split}
		& E_{r}^{\pm }=\frac{i\gamma \pm 2g_{1}^{\prime}}{2}, \\
		& \frac{c^{\pm}_{n}}{c^{\pm}_{n-1}}=\frac{-g_{2}}{-(E_{r}^{\pm
			}-i\gamma)+\nu e^{i\varphi}},
	\end{split}
\end{equation}%
where $g_1'=\sqrt{g_1^2-\gamma^2/4}$.  Eqs.\eqref{Edge}  obviously show that the edge-state eigenvalues $E_r^{\pm}$ are independent of $\nu$ and $\varphi$,  while the amplitude ratio $c_n^{\pm}/c_{n-1}^{\pm}$ depends on the NNN strength $\nu$ and the Peierls phase $\varphi$.  In other words, the loop phase primarily controls edge-state existence and localization rather than shifting the analytical edge eigenvalues.

 According to the second condition $|c_{n-1}^{\pm}|<|c_n^{\pm}|$ and Eqs.\eqref{Edge},  we can obtain an inequation as
\begin{equation}\label{Region}
	g_2^2(\theta)>g_1^2+\nu^2+\nu\gamma\sin\varphi \mp 2\nu g_1' \cos\varphi.
\end{equation}
This inequation determines the critical left and right boundaries, $\theta_{l,\pm}$ and $\theta_{r,\pm}$,  they define the stability-interval widths of the upper ($+$) and lower ($-$) edge states as $\Delta\theta_{\pm}=\theta_{r,\pm}-\theta_{l,\pm}$.

To evaluate the corresponding change in localization,  the inverse participation ratios of the upper and lower edge states can be defined as
\begin{equation}\label{IPRs} 
	\mathrm{IPRs}_{\pm}=\frac{\sum_{n}\left(
		|b^{\pm}_{n}|^{4}+|c^{\pm}_{n}|^{4}\right) }{\left[
		\sum_{n}\left( |b^{\pm}_{n}|^{2}+|c^{\pm}_{n}|^{2}\right) %
		\right] ^{2}},		
\end{equation}
 which characterizes the localization strength of the voltage distribution\cite{Helbig2020NatPhys}.  The larger values of $\mathrm{IPRs}_{\pm}$  describe the stronger localization of the edge states near the boundaries.

\begin{figure}[htbp]
	\includegraphics[width=\columnwidth]{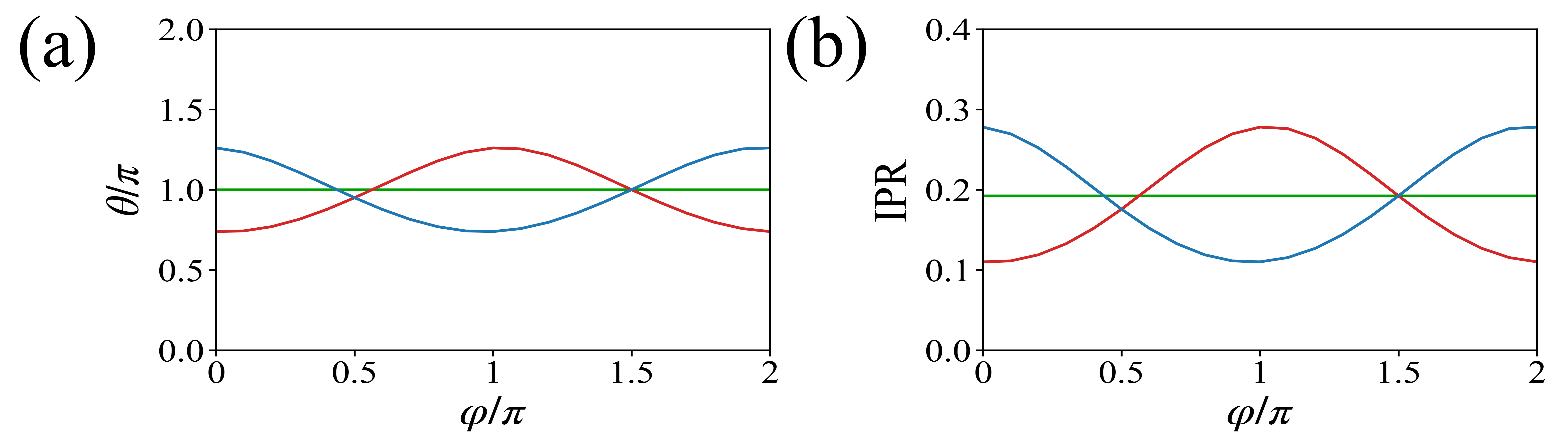}
	\caption{Peierls-phase-driven reconstruction of topological phase regions and edge-state localization. (a) The upper- and lower-edge-state stability-region widths $\Delta\theta_{\pm}$ versus the Peierls phase $\varphi$. (b) The inverse participation ratios $\mathrm{IPRs}_{\pm}$ of the upper and lower edge states  versus $\varphi$. In both panels, blue and red curves denote the upper and lower edge states, respectively, and the green horizontal line denotes the reference result for $\nu=0$. The parameters are $\gamma=0.2$ and $\nu=0.2$.}
	\label{Fig4}
\end{figure}

 According to Eqs.\eqref{Region} and \eqref{IPRs}, both $\Delta\theta_{\pm}$ and $\mathrm{IPRs}_{\pm}$ are the functions of $\varphi$ with a period $2\pi$ that are described in Figs.\ref{Fig4}(a) and (b), respectively.  Apparently, the upper and lower edge states evolve in a complementary manner. Concretely, the stability-region width and the localization strength are widened and increased in the paired edges states, inversely they are narrowed and decreased in the other ones.  As a result, the nonvanishing NNN coupling channel produces a difference in the interval widths $\varphi$ of the upper and lower edge states, and the Peierls phase can effectively manipulate the difference.

\section{Simulations and Experimental Realizations}\label{4}
 In this section, we will simulate the non-Hermitian one-dimensional trimer circuit by LTspice, and fabricate the corresponding printed circuit board (PCB).   Before the simulation and fabrication, we first consider the characteristic equation as
\begin{equation}\label{SEq}
	H_{\mathrm{circ}} V=\omega_{\pm} V,
\end{equation}
where $H_{\mathrm{circ}}$ is the effective Hamiltonian describing one cell of the trimer circuit, $\omega_{\pm}$ stand for the eigenvalues of frequencies for the upper and lower edge states, respectively  (see Appendix~\ref{B}). The response band of the upper edge state is associated with $\omega_1=1/\sqrt{(C_0+C_1+C_2)L_0}$,  and that of the lower edge state is  $\omega_2=1/\sqrt{(3C_0+C_1+C_2)L_0}$.  They provide  two channels  to observe experimentally the upper and lower edge-state responses.

\begin{figure*}[htbp]
	\includegraphics[width=\linewidth]{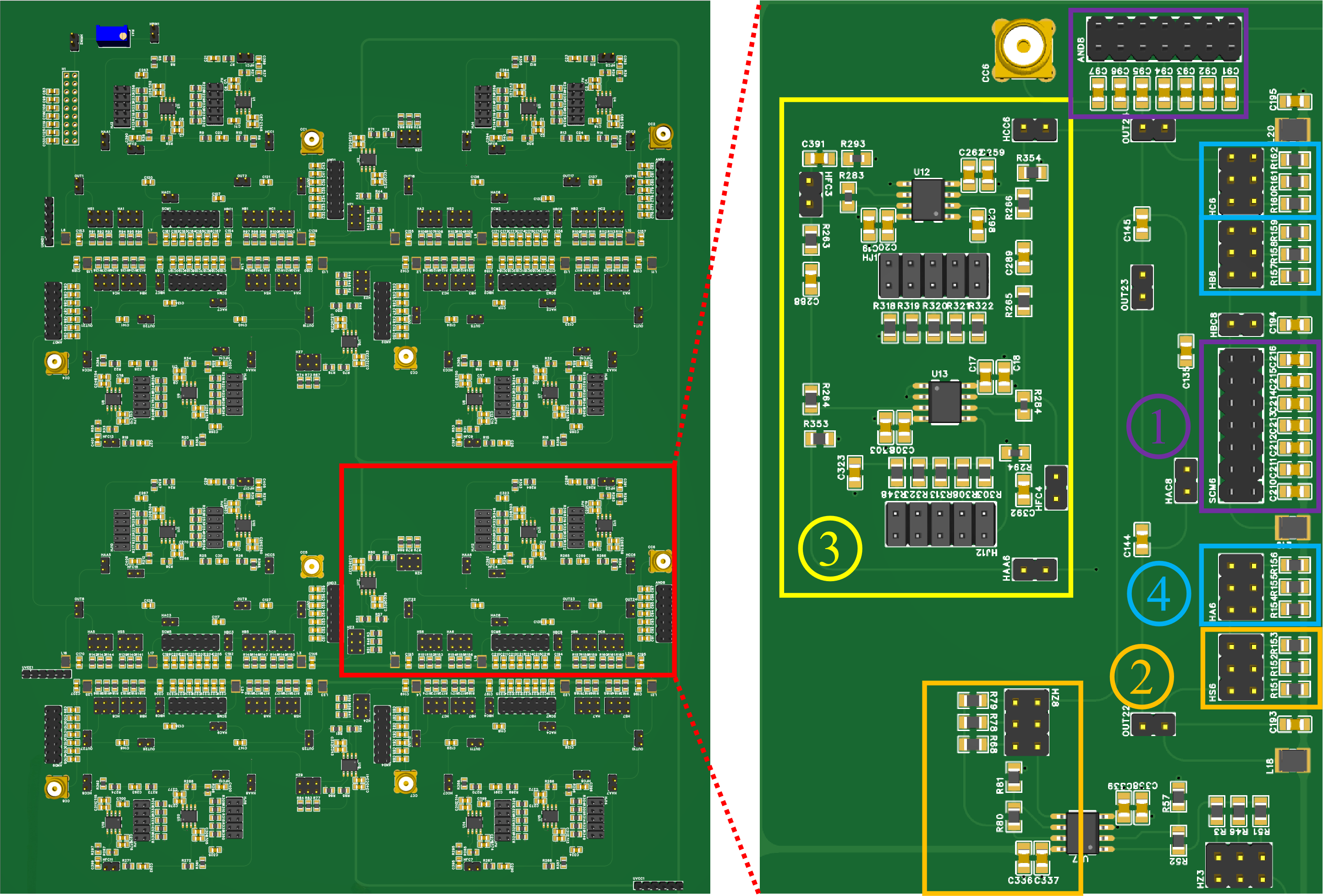}
	\caption{PCB layout of the non-Hermitian topological circuit and enlarged PCB layout of a single unit cell. The numbered labels indicate the main circuit components: (1) tunable coupling capacitor array, consisting of six $200~\mathrm{pF}$ capacitors and one $500~\mathrm{pF}$ capacitor; (2) gain/loss module, including one dual operational amplifier ADA4891-2, two $10~\mathrm{k}\Omega$ current-limiting resistors, a tunable negative-resistance module, and $100~\mathrm{nF}$ and $10~\mu\mathrm{F}$ decoupling capacitors; (3) Peierls-phase circuit module; and (4) tunable resistor $R_0$ module.}
	\label{Fig5} 
\end{figure*}

\begin{figure*}[htbp]
	\includegraphics[width=\linewidth]{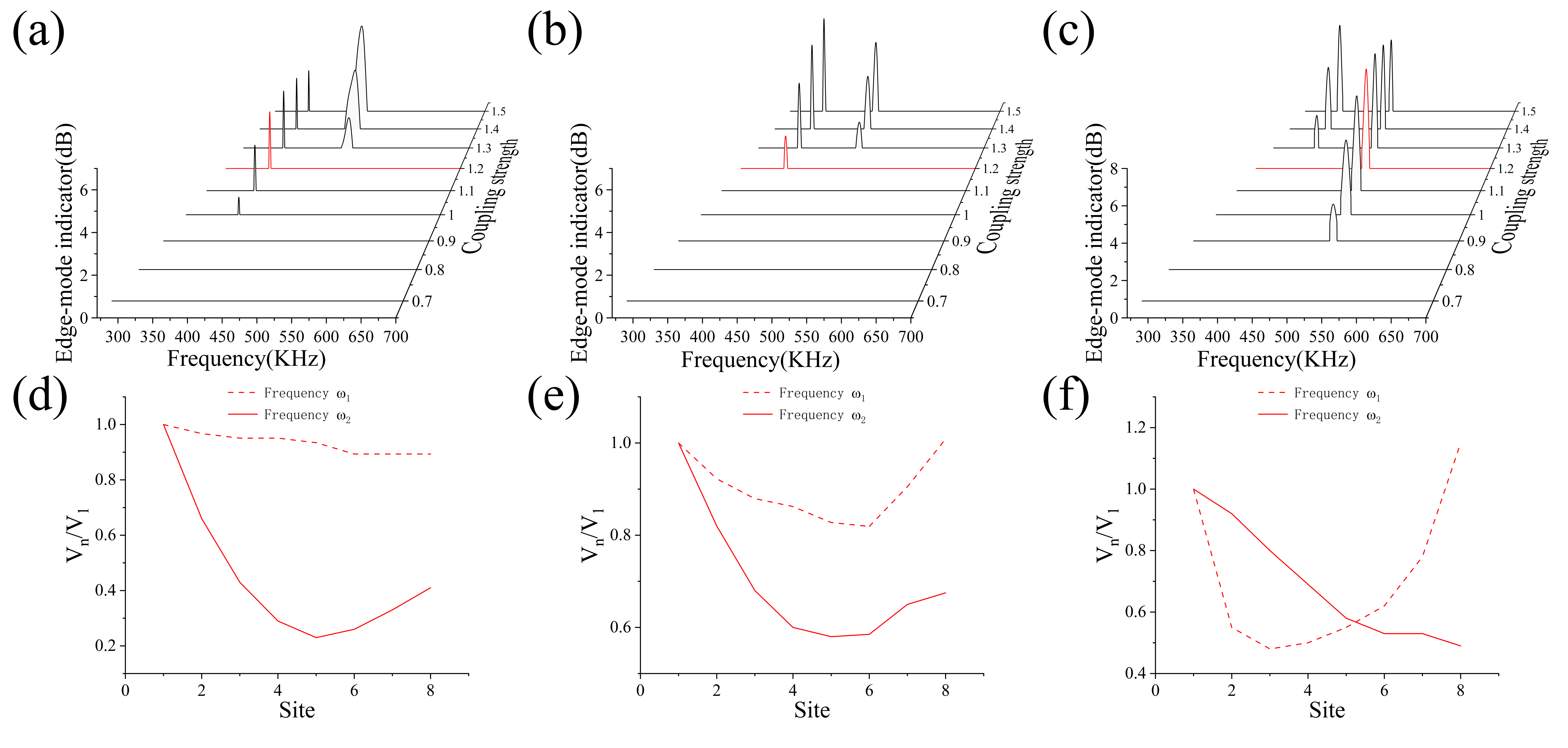}
	\caption{Edge-state indicator and experimental voltage profiles under Peierls-phase modulation. (a), (b) and (c) the distributions of edge state indicator  are described in the plane defined by the frequency $\omega$ and $g_2$ with $\nu=0.2$ for $\varphi=0^\circ$, $90^\circ$, and $180^\circ$, respectively. The red curves mark the representative line at $g_2=1.2$. (d), (e) and (f) the experimental voltage profiles $V_n/V_1$ versus lattice site at $g_2=1.2$ with $\nu=0.2$ for $\varphi=0^\circ$, $90^\circ$, and $180^\circ$, respectively. Solid and dashed lines correspond to $\omega_1$ and $\omega_2$.}
	\label{Fig6} 
\end{figure*}

By using LTspice\cite{Zou2021NatCommun}, we simulate the non-Hermitian one-dimensional trimer circuit to perform the amplitude-frequency response. In simulating process, the $C$ node in the 10th unit cell is chosen as the boundary node, and the $C$ node in the 5th unit cell is taken as the bulk one. And then, it is selected as an edge-state indicator that the amplitude-frequency response of the boundary node subtracts that of the bulk node. As shown in Figs.\ref{Fig6}(a), (b) and (c), the indicator depends on the frequency $\omega$ and the coupling parameter $g_2$ being simulated with $\nu=0.2$ at $\varphi=0^\circ$, $90^\circ$, and $180^\circ$, respectively. In the case of $\varphi=0$, the interval of the lower edge states is obviously wider than that of the upper edge states, there are more the positive-response peaks. As $\varphi=90^\circ$, the intervals are completely equivalent in the two edge states, and there are really equivalent number peaks in the upper and lower edge states, too. While $\varphi=180^\circ$, the more number peaks appear in the upper edge states. These results agree with the complementary evolution of $\Delta\theta_{\pm}$ in Fig.\ref{Fig4}(a),  which directly shows that the intervals of the upper and lower edge states can be controlled by changing the Peierls phase.

Further experimentally, we fabricate a printed circuit board (PCB) for the non-Hermitian one-dimensional trimer circuit, as shown in Fig.\ref{Fig5}. The circuit parameters  are taken as $C_0=1\,\mathrm{nF}$, $C_1=0.7\,\mathrm{nF}$-$1.5\,\mathrm{nF}$ (corresponding to $g_2=0.7$-$1.5$), and $L_0=47\,\mu\mathrm{H}$.  And then the eigenfrequencies $\omega_1=1/\sqrt{(C_0+C_1+C_2)L_0}$ and $\omega_2=1/\sqrt{(3C_0+C_1+C_2)L_0}$ can be obtained for the upper and lower edge states, respectively. In the experiment, a sinusoidal excitation signal generated by an arbitrary waveform generator (SDG2122X) is applied to the terminal node A of the circuit. To ensure identical initial excitation conditions for different measurements, the voltage amplitude at the node C of the initial unit cell is fixed at $2 \mathrm{V}$. The steady-state voltage responses at different nodes  are measured with an oscilloscope (SDS 1204X-E). Practically, the inductors have relatively low quality factors ($Q\approx40\text{--}60$), the signal therefore decays rapidly during propagation in the initial measurements. As a solution, a negative resistor is introduced to each node as compensation, enhance the effective quality factor, and  maintain the edge-state response.

For the convenience of sake, the line $g_2=1.2$ is selected as a reference that is plotted in red, as sketched in Figs.\ref{Fig6}(a), (b) and (c). And the normalized voltages of the point sites are measured for $\varphi=0^\circ, 90^\circ, 180^\circ$, as shown in Figs.\ref{Fig6}(d), (e) and (f), respectively. The spatial decay of the normalized voltage $V_n/V_1$ is employed to characterize the localization strength of the edge states. In Fig.\ref{Fig6}(d) of $\varphi=0^\circ$, the $\omega_1$ band exhibits pronounced boundary localization and decays rapidly into the bulk, and indicates a topologically nontrivial response region. However, the $\omega_2$ band decays gradually along the lattice and shows no clear boundary localization, a topologically trivial response region. In Fig.\ref{Fig6}(e) of $\varphi=90^\circ$, both the $\omega_1$ and $\omega_2$ bands show clear boundary-localization characteristics. In Fig.\ref{Fig6}(f) of $\varphi=180^\circ$, the $\omega_1$ band becomes gradually decaying and topologically trivial, whereas the $\omega_2$ band becomes strongly boundary-localized and topologically nontrivial. These results agree well with those previously discussed, which experimentally demonstrate that the response intervals of $\omega_1$ and $\omega_2$ can be regulated by changing the Peierls phase $\varphi$.

\begin{figure}[t]
	\includegraphics[width=\columnwidth]{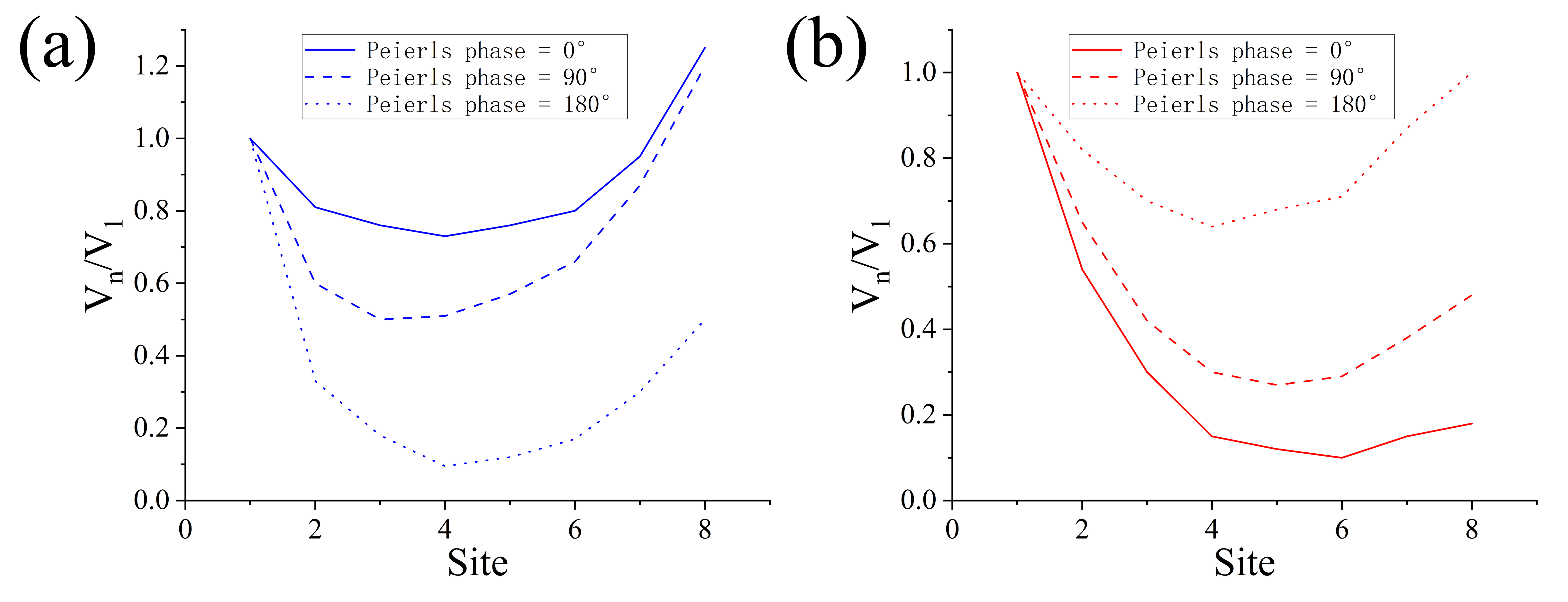}
	\caption{Experimental localization of the upper and lower edge states under Peierls-phase modulation at fixed topologically nontrivial phase. (a) Experimental voltage profiles $V_n/V_1$ versus lattice site for the upper edge state at $g_2=1.5$ with $\varphi=0^\circ$, $90^\circ$, and $180^\circ$. (b) Experimental voltage profiles $V_n/V_1$ versus lattice site for the lower edge state at $g_2=1.5$ with $\varphi=0^\circ$, $90^\circ$, and $180^\circ$. The phase-dependent voltage decay reflects the redistribution of edge-state localization strength.}
\label{Fig7} 
\end{figure}

In further examination, the previous measurements are performed again at $g_2=1.5$, which lies deeper in the nontrivial regime. In the new case, the measurements of voltage are performed for the point sites, as described in Fig.\ref{Fig7}. In Fig.\ref{Fig7}(a), the boundary voltage amplitude of the upper edge state decreases gradually as $\varphi$ increases, indicating weakened localization. The boundary voltage is largest at $\varphi=0^\circ$ and is reduced at $\varphi=180^\circ$. In Fig.\ref{Fig7}(b), the boundary voltage amplitude of the lower edge state substantially increases as $\varphi$ increases, indicating enhanced localization. These data further demonstrate that the Peierls phase produces complementary modulation of both the existence intervals and localization strengths of the two edge-state branches.

\section{Conclusions and Discussion}\label{5}
In terms of the trimer circuit of the non-Hermitian one-dimensional SSH model, we really fabricate the enhanced and reenhanced circuits by introducing the NNN coupling channels and the Peierls phase. In the initial circuit, the difference between the connection of $A$ and $C$ sites and that of $A$ and $B$ sites generates the topological upper and lower edge states, the topological phase regions present in \(0 \to 2 \to 0\). The introduced NNN coupling terms enrich the stepwise sequence as \(0 \to 1 \to 2 \to 1 \to 0\).  And the Peierls phase can provide an access to accumulate phase  around the closed loops formed by the NN and NNN paths generates an effective synthetic gauge field. This synthetic gauge field then redistributes the stability intervals and localization strengths of the upper and lower edge states.

By using numerically calculations, the \(\Delta\theta_{\pm}\) and \(\mathrm{IPR}_{\pm}\) are depending periodically on the Peierls phase. The upper and lower edge states evolve in a complementary manner. Circuit simulations and experiments further confirm that the Peierls phase shifts the response intervals of the edge states and modulates their boundary-localization characteristics. The synthetic gauge field regulates existing edge-state branches rather than creating an additional spectral framework, which  supports a programmable route to magnetic-like topological control in non-Hermitian topolectrical circuits. As a conclusion, the topological circuit is a valuable platform to investigate the topological states.

\section*{Acknowledgments}
This work is jointly supported by the National Nature Science Foundation of China (Grant No. 12475019, 12247211 and 12147103), the Natural Science Foundation of Shandong Province (Grant No. ZK2024MA038) and National Lab of Solid State Microstructure of Nanjing University (Grants No. M35040 and No. M35053).

\appendix

\section{The Circuit Equations in Schr\"{o}dinger-Like}\label{A}
In this appendix, the circuit equations in Schr\"{o}dinger-like are deduced. Generally, an RLC circuit can be described by the equation as
\begin{equation}\label{SecondKEq}
\dot{\mathbf{I}}(t) = C \ddot{\mathbf{V}}(t) + \sigma \dot{\mathbf{V}}(t) + L^{-1} \mathbf{V}(t),
\end{equation}
where $\dot{\mathbf{I}}(t)=\partial{\mathbf{I}(t)}/\partial t$ denotes the first-order time derivative of the current $\mathbf{I}(t)$, $\ddot{\mathbf{V}}(t)=\partial^2\mathbf{V}(t)/\partial t^2$ and $\dot{\mathbf{V}}(t)=\partial\mathbf{V}(t)/\partial t$ are the second- and first-order time derivatives of the electric potential $\mathbf{V}(t)$, respectively. And the matrices $L$, $\sigma$, and $C$ stand for the inductance, conductance, and capacitance of the circuit, respectively. By introducing  $\psi_{1}(t)=\dot{\mathbf{V}}(t)$ and $\psi_{2}(t)=\mathbf{V}(t)$\cite{Sahin2025APL}, the corresponding state vector could be expressed as
\begin{equation}\label{WaveFunct}
	\psi(t) = \begin{pmatrix}
		\psi_{1}(t) \\
		\psi_{2}(t)
	\end{pmatrix}
	=
	\begin{pmatrix}
		\dot{\mathbf{V}}(t) \\
		\mathbf{V}(t)
	\end{pmatrix}.
\end{equation}
With no external input current $\mathbf{I}(t)=0$ and the relation $\psi_1(t)=\dot{\psi}_2(t)$, the evolution of the state vector can be described by the following equation
\begin{equation}\label{DevEq}
	\frac{\mathrm{d}}{\mathrm{d}t}
	\begin{pmatrix}
		\psi_{1}(t) \\
		\psi_{2}(t)
	\end{pmatrix}
	=
	\begin{pmatrix}
		-C^{-1}\sigma & -C^{-1}L^{-1} \\
		\mathbb{E} & 0
	\end{pmatrix}
	\begin{pmatrix}
		\psi_{1}(t) \\
		\psi_{2}(t)
	\end{pmatrix},
\end{equation}
where $\mathbb{E}$ is an identity matrix. Multiplying two sides of Eq.\eqref{DevEq} by $-i$ the above derivative equation can be rewritten into a Schr\"{o}dinger-like form $-i\,\mathrm{d}\psi(t)/\mathrm{d}t=H\psi(t)$, where the effective Hamiltonian is
\begin{equation}\label{EffHam}
	H=i
	\begin{pmatrix}
		C^{-1}\sigma & C^{-1}L^{-1} \\
		-E & 0
	\end{pmatrix}.
\end{equation}
Here, $H$ is the effective time-domain Hamiltonian of the circuit; its eigenvalues determine the resonant frequencies, and its eigenvectors describe the corresponding voltage distributions. This construction connects the frequency-domain Laplacian $J_{ab}(\omega)$ to the time-domain evolution governed by $H$.

\section{Derivation of the Eigenfrequencies of Upper and Lower Edge States}\label{B}
In this appendix, the eigenfrequencies are deduced for the two edge-state response channels. For the circuit containing the NNN coupling capacitor $C_2$, according to the Kirchhoff's current law, a trimerized unit cell with the nodes $A$, $B$ and $C$ can be described by
\begingroup
\footnotesize
\begin{equation}
	\left\{
	\begin{aligned}
		I_A &= \left(\frac{1}{R}+\frac{1}{R_0}+i\omega C_0+\frac{1}{i\omega L_0}\right)V_A
		+i\omega C_0(V_A-V_B)\\
		&\quad +i\omega C_1(V_A-V_Ce^{-ik})
		+i\omega C_2(V_A-V_C),\\
		I_B &= \left(\frac{1}{R_0}+i\omega C_1+i\omega C_2+\frac{1}{i\omega L_0}\right)V_B
		+i\omega C_0(V_B-V_A)\\
		&\quad +i\omega C_0(V_B-V_C),\\
		I_C &= \left(-\frac{1}{R}+\frac{1}{R_0}+i\omega C_0+\frac{1}{i\omega L_0}\right)V_C
		+i\omega C_0(V_C-V_B)\\
		&\quad +i\omega C_1(V_C-V_Ae^{ik})
		+i\omega C_2(V_C-V_A).
	\end{aligned}
	\right.
\end{equation}
\endgroup
For eigenmodes, the external currents vanish, namely $I_A=I_B=I_C=0$. The above circuit equation can be simplified in matrix form as
\begin{equation}
	J(\omega,k)
	\begin{pmatrix}
		V_A\\
		V_B\\
		V_C
	\end{pmatrix}
	=0,
\end{equation}
with
\begin{equation}
	\begin{aligned}
		J(\omega,k) &= i\omega \left( 2C_0 + C_1 + C_2 - \frac{1}{\omega^2 L_0} \right) E \\
		&\quad + i\omega
		\begin{pmatrix}
			\frac{\sigma+\sigma_0}{i\omega} & -C_0 & -C_1 e^{-ik}-C_2 \\
			-C_0 & \frac{\sigma_0}{i\omega} & -C_0 \\
			-C_1 e^{ik}-C_2 & -C_0 & \frac{-\sigma+\sigma_0}{i\omega}
		\end{pmatrix}.
	\end{aligned}
\end{equation}

By normalizing with $i\omega C_0$, the above equation can be further simplified as
\begin{equation}
	J_n(\omega,k)\mathbf{V}=0,
\end{equation}
where $\mathbf{V}=(V_A,V_B,V_C)^T$ and
\begin{equation}
	\begin{aligned}
	J_n(\omega,k)
	&=
	\left[
	2+g_2+\nu-\frac{1}{\omega^2 L_0 C_0}
	+\frac{1}{i\omega R_0 C_0}
	\right]\mathbb{E}\\
	&\quad +Y(\omega,k),
	\end{aligned}
\end{equation}
where $\mathbb{E}$ is an identity matrix, and
\begin{equation}
	Y(\omega,k)=
	\begin{pmatrix}
		\frac{1}{i\omega R C_0} & -1 & -g_2 e^{-ik}-\nu \\
		-1 & 0 & -1 \\
		-g_2 e^{ik}-\nu & -1 & -\frac{1}{i\omega R C_0}
	\end{pmatrix}.
\end{equation}
Near the two edge-state resonances, the gain-loss coefficient can be approximately written as
\begin{equation}
	\frac{1}{i\omega R C_0}\simeq -i\gamma.
\end{equation}
The Peierls phase is introduced in the NNN branch, where it adds opposite complex phases to the two directions. The  matrix $Y(\omega, k)$ will be rewritten as
\begin{equation}
	Y(\omega,k)=
	\begin{pmatrix}
		-i\gamma & -1 & -g_2 e^{-ik}-\nu e^{i\varphi} \\
		-1 & 0 & -1 \\
		-g_2 e^{ik}-\nu e^{-i\varphi} & -1 & i\gamma
	\end{pmatrix}.
\end{equation}

The eigenmode condition $J_n(\omega,k)\mathbf{V}=0$ is equivalent to the secular equation
\begin{equation}
	Y(\omega,k)\mathbf{V}
	=
	-\left[
	2+g_2+\nu-\frac{1}{\omega^2 L_0 C_0}
	+\frac{1}{i\omega R_0 C_0}
	\right]\mathbf{V},
\end{equation}
where the resistor $R_0$ acts as a uniform loss offset and compensates the imaginary part of the edge-state eigenvalue. Therefore, the real part of the characteristic frequency is determined by
\begin{equation}
	2+g_2+\nu-\frac{1}{\omega^2 L_0 C_0}
	+\lambda=0,
\end{equation}
where $\lambda$ denotes the real eigenvalue with respect to the edge-state channel of $Y(\omega,k)$.

For the upper edge-state channel, the  eigenvalue is approximated by
\begin{equation}
	Y(\omega,k)\mathbf{V}_1\simeq -\mathbf{V}_1.
\end{equation}
Therefore, $\lambda_1=-1$, and the characteristic frequency is obtained from
\begin{equation}
	2+g_2+\nu-\frac{1}{\omega_1^2 L_0 C_0}-1=0,
\end{equation}
and then the eigenfrequency $\omega_1$ can be obtained as
\begin{equation}
	\omega_1
	=
	\frac{1}{\sqrt{(1+g_2+\nu)C_0L_0}}
	=
	\frac{1}{\sqrt{(C_0+C_1+C_2)L_0}}.
\end{equation}

For the lower edge-state channel, the approximation is
\begin{equation}
	Y(\omega,k)\mathbf{V}_2\simeq \mathbf{V}_2.
\end{equation}
And then $\lambda_2=+1$, and the characteristic frequency is determined by
\begin{equation}
	2+g_2+\nu-\frac{1}{\omega_2^2 L_0 C_0}+1=0,
\end{equation}
as
\begin{equation}
	\omega_2
	=
	\frac{1}{\sqrt{(3+g_2+\nu)C_0L_0}}
	=
	\frac{1}{\sqrt{(3C_0+C_1+C_2)L_0}}.
\end{equation}

The two frequencies provide the analytical assignment of the edge-state response peaks observed in the circuit spectrum.

\bibliographystyle{apsrev4-1}
\bibliography{circuit}

\end{document}